\documentclass[journal]{IEEEtran}
\usepackage{datetime}
\usepackage{times}
\usepackage{epsfig,graphicx}  
\usepackage{graphicx, amssymb}
\usepackage{amsfonts, enumerate}
\usepackage{latexsym,subfigure,makeidx}
\usepackage{url}
\usepackage{tabls}
\usepackage{algorithm}
\usepackage{algpseudocode}
\usepackage{lscape}
\usepackage[sort,space]{cite}
\usepackage{pifont}
\usepackage{tabularx}
\usepackage{rotating}
\usepackage{hhline}
\usepackage{textcomp,booktabs}
\usepackage{xcolor}
\usepackage{array}
\usepackage{amsmath,bm}
\usepackage{caption}
\usepackage{multirow}
\usepackage{verbatim}
\usepackage{amsthm}
\usepackage{framed}

\ifCLASSINFOpdf
\else
\fi
\ifCLASSOPTIONcompsoc
  \usepackage[caption=false,font=normalsize,labelfont=sf,textfont=sf]{subfig}
\else
 \usepackage[caption=false,font=footnotesize]{subfig}
\fi
\def\NoNumber#1{{\def\alglinenumber##1{}\State #1}\addtocounter{ALG@line}{-1}}
\begin{document}
%
\title{Deep Reinforcement Learning and Permissioned Blockchain for Content Caching in Vehicular Edge Computing and Networks}
%
%

%

\author{Yueyue~Dai, 
        Du~Xu, 
        Ke Zhang,
         Sabita~Maharjan,~\IEEEmembership{Senior~Member,~IEEE}
        and~Yan~Zhang,~\IEEEmembership{Fellow,~IEEE}
\thanks{This research is partially supported by the Key R\&D Project of Sichuan Province, under Grant No. 2019YFG0520, and the Opening Project of Shanghai Trusted Industrial Control Platform, under Grant No. 2019YY201.}        
\thanks{Y. Dai and D. Xu  and  K. Zhang are with the  University of Electronic Science and Technology of
China, Chengdu, China (email:yueyuedai@ieee.org; xudu.uestc@gmail.com; zhangke@uestc.edu.cn ).}
\thanks{Sabita Maharjan is with the Simula Metropolitan Center for Digital
Engineering, 0167 Oslo, Norway, and also with the University of Oslo, 0316
Oslo, Norway (e-mail: sabita@simula.no). }
\thanks{Y.  Zhang (corresponding author)  is with the Department of Informatics, University of Oslo, 0316
Oslo, Norway, and also  with Simula Metropolitan Center for Digital Engineering, Norway.(email: yanzhang@ieee.org).}
}

\maketitle

\begin{abstract}
Vehicular Edge Computing (VEC) is a promising paradigm  to enable huge amount of data and multimedia content  to be cached in proximity to vehicles. However,  high mobility of vehicles  and dynamic wireless channel condition make it challenge to  design an optimal content caching policy. Further, with much sensitive personal information, vehicles may be not willing to caching their contents to an untrusted caching provider. Deep Reinforcement Learning (DRL) is an emerging technique to solve the problem with high-dimensional and time-varying features. Permission blockchain  is able to  establish a secure and decentralized peer-to-peer transaction environment. In this paper, we integrate DRL and permissioned blockchain  into vehicular networks for intelligent and secure  content caching. We first propose a  blockchain empowered distributed  content caching framework  where vehicles perform  content caching and base stations maintain the permissioned blockchain. Then, we exploit the advanced DRL approach to  design an optimal content caching scheme with taking mobility into account.  Finally, we propose  a new block verifier selection method,  Proof-of-Utility (PoU), to accelerate block verification process. Security analysis shows that our proposed blockchain empowered content caching can achieve security and privacy protection. Numerical results based on a real dataset from Uber indicate that the DRL-inspired content caching scheme significantly outperforms two benchmark policies.
\end{abstract}

\begin{IEEEkeywords}
 Deep Reinforcement Learning, Permissioned Blockchain, Content Caching, Vehicular Edge Computing
\end{IEEEkeywords}

%
\IEEEpeerreviewmaketitle

\section{Introduction}

With the rapid development of in-car touchscreen and autonomous driving systems (e.g., Tesla Autopilot),  huge amount of data and  content generated by in-vehicle sensors and  vehicular infotainment applications \cite{8403956}, \cite{8264740}. However, long distance between vehicles and cloud servers, and the limited backhaul link capacity pose significant challenges for supporting massive content delivery while also satisfying the low-latency requirement in vehicular networks \cite{7995077}.  Vehicular Edge Computing (VEC) is a promising paradigm where Base Stations (BS) and vehicles with a certain amount of computation resource and caching resource can be utilized as edge servers to cooperatively cache content at the network edge\cite{8322166},\cite{7636965},\cite{dai2018joint},\cite{8667693}. 


Caching content at edge servers  can  effectively alleviate mobile traffic on backhaul links and reduce content delivery latency \cite{8758209}.  The authors in \cite{shanmugam2013femtocaching} proposed to cache content on femto-cell base stations  to minimize the total expected content delivery delay based on a given popularity distribution. 
Since state-of-the-art vehicles are equipped with a certain amount of caching resource, the vehicle with sufficient caching resource can be regarded as a caching provider to expand the caching capacity of the network edge.  Vehicle-to-vehicle communication can further reduce average content transmission latency  \cite{7166190},\cite{6807947}, \cite{8039268}.  However, high mobility of vehicles leads to dynamic network topology and  time-varying wireless channel condition which makes it difficult to design an optimal content caching policy\cite{6970763}, \cite{8447267}.  Moreover,  a content usually involves much sensitive personal information of its generator such that vehicles may be not willing to store their contents to an untrusted caching provider. 

Deep Reinforcement Learning (DRL) is an emerging technique which has the ability to learn and build knowledge about dynamic wireless communication environment  \cite{dai2018AI}. By interacting with edge servers, the authors in \cite{dai2018AI}  utilized DRL to observe the available computing and caching resource at the network edge and design the corresponding resource allocation scheme.  Exploiting actor-critic reinforcement learning, the authors in \cite{wei2018joint} proposed a scheme to solve the joint content caching, computation offloading, and resource allocation problems in fog-enabled Internet of Things (IoT) networks. The authors in \cite{zhang2019deep} proposed a deep Q-learning based task offloading scheme to select an optimal edge server for vehicles to maximize task offloading utility. However, security and privacy are not considered in the above  works.

 Blockchain is an open database which maintains an immutably distributed ledger to enable securely transactions among distributed entities without relying on a central intermediary \cite{dai2019blockchain}, \cite{8579189},\cite{8843900}. Blockchain can be categorized into two main types: public blockchain and permissioned blockchain.  In public blockchain,  anyone can participate in the process of verifying transactions and creating blocks due to no access limitation, such as Bitcoin and Ethereum.  In permissioned blockchain, only permissioned nodes can verify transactions and create blocks. {The typical consensus in public blockchain is  computation-intensive  because nodes compete against each other for  creating newly blocks by solving a difficult PoW puzzle.  Because of no competitive PoW puzzle, permissioned blockchain  can build a distributed ledger with less energy and computation resource.  On the other hand,  in public blockchain, all distributed nodes have to participate in the process of consensus which results in a long time consumption.  In contrast, in permissioned blockchain, the number of nodes participating in the consensus is quite few, such that this type of blockchain can achieve a very fast consensus. Thus, permissioned blockchain is suitable for energy-constrained and delay-sensitive networks. Because of  the limited energy and computation resources of vehicles and the stringent delay requirement of applications, the existing work often select permissioned blockchain with low energy consumption and short consensus delay  for vehicular networks   \cite{8579189}, \cite{8734799},\cite{LUTVT},\cite{kang2019reliable}.  }

In this paper, we integrate DRL and permissioned blockchain  into vehicular networks to propose an intelligent and secure content caching scheme. We first propose a  blockchain empowered distributed and secure content caching framework  where vehicles act as caching requesters and caching providers to perform content caching and BSs act as verifiers to  maintain permissioned blockchain. Due to high mobility of vehicles, we exploit the advanced DRL approach to learn dynamic network topology and time-varying wireless channel condition and then design an optimal content caching scheme between caching requesters and caching providers. We utilized permissioned blockchain to ensure a secure  content caching among vehicles. To enable a fast and efficient  blockchain consensus mechanism, we propose to select block verifiers based on Proof of Utility (PoU).  The main contributions of this paper are summarized as follows:
\begin{itemize}
\item  We propose a  blockchain empowered distributed and secure content caching framework  where vehicles perform content caching and  BSs  maintain permissioned blockchain to ensure an intelligent and secure  content caching.

\item We formulate the content caching problem as the form of DRL  to maximize content caching with taking vehicular mobility into account and  design a new DRL-inspired content caching scheme.

\item We design  a new block verifier selection method to enable a fast and efficient  blockchain consensus mechanism. Security analysis shows that our proposed blockchain empowered content caching can achieve security and privacy protection. Numerical results based on a real dataset demonstrate the effectiveness of the proposed DRL-inspired content caching scheme. 
\end{itemize}

The remainder of this paper is organized as follows. We introduce the architecture of blockchain empowered content caching in  Section  \ref{sm}. Then, we propose a DRL-inspired content caching scheme and introduce PoU consensus in Section \ref{sectionii} and Section \ref{blockchain}, respectively. We present the security analysis and numerical results  in Section \ref{result}. Finally, we conclude this paper  in Section \ref{c}.

\section{Related Work}

{
Recently, blockchain technology  has attracted enormous attention of researchers and developers because of  its feature such as decentralization, immutability, anonymity, and security. The authors in \cite{8734799} proposed a neural-blockchain based drone-caching approach in  unmanned aerial vehicles where  blockchain ensures the high reliable communication among drones.   The authors in \cite{8422547} proposed a blockchain-based proactive caching in hierarchical wireless networks to  enable autonomous caching-delivery among untrustworthy parties.  The authors in \cite{8358773}   proposed a decentralized data management scheme for vehicular networks based on  blockchain. However, these works  establish blockchain  by solving the meaningless PoW puzzle or  Proof-of-Stake (PoS). The characteristics of nodes in vehicular edge computing networks, such as computing ability or QoS requirement, are not involved in the process of  blockchain establishment.}

To make a better integration  of blockchain and vehicular networks, a few studies have utilized PoX consensus to replace the original PoW schemes. {  Because of no competitive PoW puzzle, the authors in \cite{7795984} indicated that Delegated Proof-of Stake (DPoS) is particularly suitable for lightweight vehicles to establish blockchain-based  transportation systems. As a further exploration of \cite{7795984}, the  authors in \cite{8624307} proposed an enhanced DPoS consensus for a  blockchain-based  vehicular data sharing system, where reputation is used in the DPoS to measure the quality of RSUs.} The authors in \cite{kang2019reliable} and \cite{8832210} also utilized reputation-based  consensus to build  blockchain for vehicular  networks or cellular networks. The authors in \cite{singh2017blockchain} utilized proof-of-driving based blockchain to enable an intelligent vehicular data sharing among vehicles.  The authors in \cite{8493118}  utilized proof-of-integrity in vehicular blockchain to manage the collected vehicle-related data  from hundreds of sensors  for a privacy-aware traffic accident diagnosis. However, the above researches are not suitable for vehicular edge computing networks, which are still not considering the limited computation resource of edge nodes and the stringent latency requirement s of users. In this paper, we proposed a proof-of-utility based consensus for vehicular edge caching, where the utility is comprehensive function to measure the computing and processing abilities of edge nodes and  the latency requirements of vehicles.

\section{Blockchain and Artificial Intelligence Content Caching}
\label{sm}

\begin{figure}
	\centering
	\includegraphics[width =3.5in]{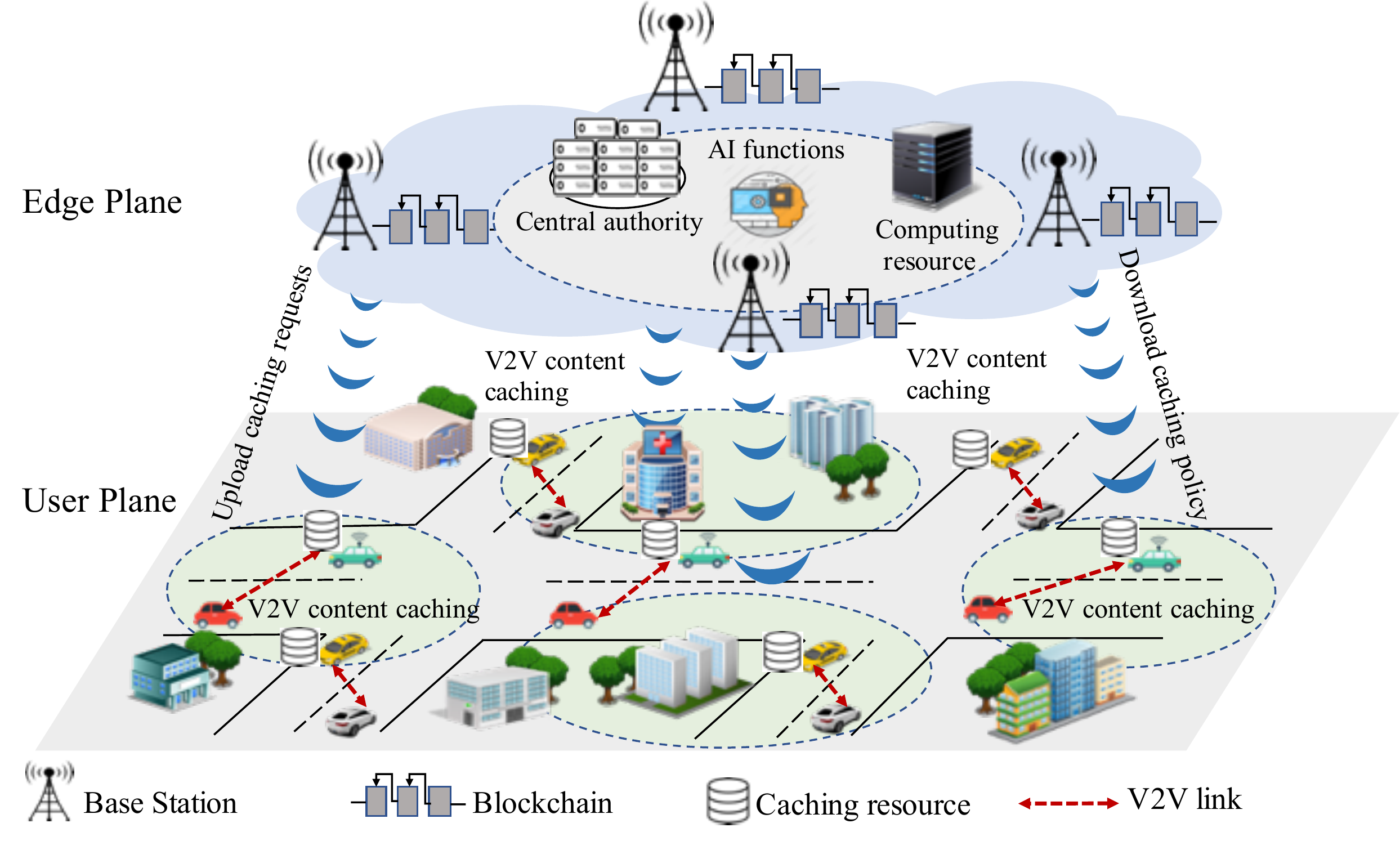}
	\caption{Blockchain empowered vehicular content caching}
	\label{overview}
\end{figure}

In this section, we first present  the proposed  blockchain empowered vehicular content caching architecture, and then describe the detailed phases of the proposed blockchain empowered vehicular content caching.

\subsection{Architecture of Vehicular Content Caching  with Blockchain}

{We propose a new blockchain empowered  content caching architecture which consists of a user plane and an edge plane, as illustrated in Fig. \ref{overview}. }

 In the user plane, vehicles, equipped with multiple sensors and applications, can collect a variety of valuable content about vehicles, roads and their surrounds, such as  entertainment videos, road maintenance information, parking lot occupancy and so on. Since state-of-the-art  vehicles have a certain amount of caching resource, they can cache their content locally. However, due to the capacity limitation of caching resource, when a resource-constrained vehicle cannot store its collected content on its own cache, anyone of its neighbor can act as a caching provider to offer its unoccupied caching resource for content caching via Vehicle-to-Vehicle (V2V) communication.  To encourage vehicles to contribute their unused caching resource, we utilize incentive mechanisms to motivate vehicles to participate in V2V content caching.
 
 In the edge plane, several BSs are distributed in a specific area to work as edge servers with communication, computing capability, and AI functions. BSs can detect  available caching resource of vehicles and  deliver  caching requests to the caching provider. In addition, BS can utilize computing capabilities and AI functions to predict the V2V  transmission range and connection duration between caching requesters and caching providers, and  perform caching pair matching to enhance system utility.  There exists a central authority in the edge plane to manage the security parameters and keys of BSs and vehicles with a  tamper-resistant hardware.

 Caching content at vehicles can enhance spectrum utilization and reduce average content delivery latency between vehicles.  However, since a content involves much sensitive and critical personal information of its generator, caching requesters are not willing to store their content to untrusted caching providers. To cope with this, each BS is equipped with a blockchain  to enable  untrustworthy vehicles to interact with each other for content caching in a secure manner.

\subsection{Blockchain-based Vehicular Content Caching}
\label{block_overview}
In Fig. \ref{overview}, there are two types of vehicles driving on the road.  We define the vehicle requiring caching resource to store its content as caching requester and  define the vehicle to provide caching resource as  caching provider.  Based on blockchain, a secure vehicular content caching can be achieved through the following phases.


1) \textbf{Identity Establishment and System Initialization:} 

To implement V2V content caching, vehicles  should register unique accounts and create their keys firstly. We utilize elliptic curve digital signature algorithm and asymmetric cryptography to establish identity. Specifically, vehicles and BSs register a legitimate identity after passing the authentication of the central authority. The legitimate identity consists of a public key, a private key, and a certificate, which can be described as $\{PK_{v_i},SK_{v_i}, Cert_{v_i} \}$.  The public key is regarded as the source address of the caching transaction which is used to verify the genuineness of  transactions.  The cryptographic private key is used to sign a transaction and  the certificate is to uniquely identify the vehicle through binding registration information of the vehicle.


 
 Each vehicle has a wallet. The wallet address is generated from its public key.  At the system initialization stage, each vehicle  requests the wallet addresses of other vehicles  from the central authority. Specifically, each vehicle uploads its wallet address to a global  account pool and then downloads other vehicles'  wallet address for content caching from there. Note that vehicles can use changeable wallet address to preserve anonymity and privacy.



2) \textbf{Triggering  Content Caching Smart Contract:}

 For content caching, each BS gathers all vehicles' caching requests and monitors their available caching resource under its coverage.  
Vehicle $v_i$ sends its caching request  to the nearest BS  $b_j$. The  caching request message of  vehicle $v_i$   includes the required caching resource $c_{v_i}$,  current location  $loc_{v_i}$, public key $PK_{v_i}$, signature $Sig_{v_i}$, certificate $Cert_{v_i}$, and timestamp $ts$, which can be described as 
\begin{equation}
Req^{v_i\rightarrow b_j} = E_{PK_{b_j}}(c_{v_i}||loc_{v_i}|| PK_{v_i}|| Sig_{v_i}||Cert_{v_i}||ts),
\end{equation}
{where $E_{PK_{b_j}}$ denotes that message $Req^{v_i\rightarrow b_j} $ is encrypted with public key $PK_{b_j}$, $Sig_{v_i} = Sign_{SK_{v_i}} ( c_{v_i}||loc_{v_i})$ denotes that the digital signature of $c_{v_i}$ and $loc_{v_i}$  is with private key $SK_{v_i}$, and  $ts$ is the timestamp of the current message.


Vehicle $v_p$ periodically sends its available caching resource  to the nearest BS  $b_j$ for caching  resource sharing.  The  message about available caching resource includes the available caching resource $C_{v_p}$, location $loc_{v_p}$, public key $PK_{v_p}$, signature $Sig_{v_p}$, certificate $Cert_{v_p}$, and timestamp $ts$,  which can be described as 
\begin{equation}
Mes^{v_p\rightarrow b_j} = E_{PK_{b_j}}(C_{v_p}||loc_{v_p}|| PK_{v_p}|| Sig_{v_p}  ||Cert_{v_p}||ts),
\end{equation}
where $E_{PK_{b_j}}$ denotes that message $Mes^{v_p\rightarrow b_j} $ is encrypted with public key $PK_{b_j}$, $Sig_{v_p} = Sign_{SK_{v_p}} (C_{v_p}||loc_{v_p})$ denotes that the digital signature of $C_{v_p}$ and $loc_{v_p}$  is with private key $SK_{v_p}$, and  $ts$ is the  timestamp of the current message.

After receiving  caching requests and available caching resource of vehicles,  BSs first verify their identity. { To speed up the verification process, BSs  adopt batch verification process which can verify the validity of a number of identities simultaneously \cite{camenisch2012batch}. Specifically, each BS abstracts the verification parameters from the received $Req^{v_i\rightarrow b_j} $ and $Mes^{v_p\rightarrow b_j} $ and constructs the verification parameters as $<PK,Mes,Sig>$, where $Mes$ is the detailed message.  Then, BS calls the batch verification algorithm for identity verification. If all $Ver(PK_{v_i},Mes_{v_i},Sig_{v_i})$ $ =1$ for all $i\in \mathcal{I} \cup \mathcal{P}$, we have $Batch ((PK_{v_1},Mes_{v_1},Sig_{v_1})$ $,...,(PK_{v_i},Mes_{v_i},Sig_{v_i}),...) =1$ and batch verification is passed.  If one or more than one of  signatures are invalid, batch verification fails.} After the batch verification, BSs  perform V2V content caching mechanism to make caching pair matching. More details on vehicular content caching  will be given in Section \ref{sectionii}.


When vehicular content caching mechanism is completed,  each BS responses a message to caching requester and caching provider,  respectively.  $Resp_{req}^{b_j\rightarrow v_i}$  is the  message that BS  $b_j$ responsing to  caching requester $v_i$ and $Resp_{pro}^{b_j\rightarrow v_r}$ is  the  message that BS  $b_j$ responsing to caching provider $v_p$, which  can respectively be described as 
\begin{equation}
\begin{aligned}
& Resp_{req}^{b_j\rightarrow v_i}= E_{PK_{v_i}}(loc_{v_p}|| chan_{ip}|| PK_{v_p}||Sig_{b_j} ||ts),\\
&Resp_{pro}^{b_j\rightarrow v_p}= E_{PK_{v_p}}(c_{v_i}||loc_{v_p}|| chan_{ip}|| Sig_{b_j} ||ts),
\end{aligned}
\end{equation}
where $chan_{ip}$ is the wireless channel between the caching requester and the caching provider.  $E_{PK_{v_i}}$ and $E_{PK_{v_p}}$ denote that  $ Resp_{req}^{b_j\rightarrow v_i}$ and $Resp_{pro}^{b_j\rightarrow v_p}$ are encrypted with public key $PK_{v_i}$ and $PK_{v_p}$, respectively. The digital signature in $Resp_{req}^{b_j\rightarrow v_i}$ is denoted as $Sig_{b_j} = Sign_{SK_{b_j}} (loc_{v_p}|| chan_{ip})$, and the digital signature in $Resp_{pro}^{b_j\rightarrow v_p}$ is denoted as $Sig_{b_j} = Sign_{SK_{b_j}} (c_{v_i}||loc_{v_p}|| chan_{ip})$.

Based on the above two messages, vehicles autonomous  execute the pre-programmed smart contract. The smart contract consists of two modules. The first module is to carry out  content delivery. The second module is to transfer a certain amount of coins from the wallet of  content caching requester $v_i$  to the wallet of  content caching provider vehicle $v_p$.

3) \textbf{Recording Transactions:}

After finished content caching,  caching requesters pays for its caching provider and generates a transaction to record the caching event. Specifically, caching requester $v_i$ sends the generated transaction  to the nearest BS $b_j$. The BS first verifies the received transaction, and then encrypts and broadcasts it to the entire blockchain network. The transaction  includes the shared caching resource $c_{v_i}$, the coins that caching provider  $v_p$ obtains $coin^{v_i\rightarrow v_j}$, the  wallet addresses of  the content requester and the content provider, the signature of the content requester,  and timestamp, namely,
\begin{equation}
\begin{aligned}
Trans^{v_i\rightarrow b_j} =& E_{PK_{b_j}}(c_{v_i}||coin^{v_i\rightarrow v_p}|| wallet_{addr}^{v_i}||\\&wallet_{addr}^{v_p}||Sig_{v_i}||ts),
\end{aligned}
\end{equation}
where $E_{PK_{b_j}}$ denotes that $Trans^{v_i\rightarrow b_j} $ is encrypted with public key $PK_{b_j}$, $Sig_{v_i} = Sign_{SK_{v_i}} ( c_{v_i}||coin^{v_i\rightarrow v_p} )$ denotes that the digital signature of $c_{v_i}$ and transferred coins  $coin^{v_i\rightarrow v_j}$  with private key $SK_{v_i}$.
The new generated transaction is broadcasted over the entire network for audit and verification. 


The verified transactions are ordered and batched into a cryptographically tamper-evident data structure, named block.  The blocks are linked in a linear chronological order by hash pointers to form a blockchain.  

4) \textbf{Building Block and Performing Consensus Process:}

Each block is  created by a specific BS  in the consensus process.  We define the BS to create the newly block as the leader. After the newly block created,  the leader broadcasts  the block with timestamp  for block audit and verification.  The other BSs   verify the correctness of the newly created block. According to Bitcoin, the fastest node which solves  Proof-of-Work (PoW) puzzle becomes the leader to create the newly block. However,  PoW puzzle is a computation-intensive and energy-consuming task such that it is not suitable for vehicular networks \cite{dai2019blockchain}. Therefore, we need a  fast and efficient  blockchain consensus mechanism with low energy-consuming  and time-consuming.

   In this paper, we aim to  design an  intelligent and secure vehicular content caching scheme for the proposed architecture. However, there are two challenges to achieve this. One is in the user plane that high mobility of vehicles leads to dynamic network topology and  time-varying wireless channel condition making it difficult to design an optimal content caching policy. The other is in the edge plane that  how to achieve fast  permission blockchain with low  energy-consuming  and time-consuming. To address such issues, we propose a DRL-inspired content caching scheme  in Section \ref{sectionii} and PoU consensus mechanism for permissioned blockchain  in Section \ref{blockchain}.






\section{Deep Reinforcement Learning-based Vehicular Content Caching }
\label{sectionii}

In this section, we propose a DRL-inspired content caching algorithm with taking vehicular mobility into account to solve the challenge in the user plane. 

\subsection{Content Caching with Manhattan grid mobility model}

We formulate  V2V content caching problem to maximize system utility by focusing on a single cell with a BS and $N$ vehicles. The BS can communicate with any  vehicle under its coverage.  We denote the set of  caching requester as $\mathcal{I} = \{v_1,...,v_I\}$ and the set of  caching provider as  $\mathcal{P} = \{v_1,...,v_P\}$, where $\mathcal{I} \cap\mathcal{P} = \varnothing  $ and $I+P = N$. The content generated by caching requester  $v_i$ can be described as $\{c_{v_i}, \tau_{v_i}\}$, where $ c_{v_i}$ and $\tau_{v_i}$ denote the required caching resource and the maximal content delivery latency,  respectively. 

{We model the city as a Manhattan style grid, with a uniform block size across a fixed square area. The Manhattan grid model is  introduced as a standard  mobility  by the European Telecommunications Standards Institute (ETSI) \cite{umts1998101}. In Manhattan grid model,  the map is composed of a number of horizontal and vertical streets. Each street has two lane for each direction (i.e., north and south direction for vertical streets, and east and west for horizontal streets). Vehicles move along streets and may turn at cross streets (i.e., intersection) with a given probability.  Let $\eta$ denote the driving direction of vehicles, where  $\eta \in\{north, south, west, east\}$. The probability that each vehicle moves at an intersection can be denoted as
\begin{equation}
\label{move}
\mathbb{P}_{\eta}  =\frac{\frac{1}{\delta_{int}\nu}}{\frac{1}{\delta_{int}\nu}+\frac{T^{wait}P^{wait}}{2}}=\frac{2}{2+T^{wait}P^{wait}\delta_{int}\nu}, 
\end{equation}
where $\delta_{int}$ is the density of intersections, $\nu$ denotes constant velocity, $T^{wait}$ denotes the maximum tolerant waiting time of vehicles at the intersection, $P^{wait}$ is the probability that vehicles  have to wait. 
Further, vehicles may stop at an intersection, which can be denoted as $\zeta$.  The probability that a vehicle stops at an intersection is represented as
\begin{equation}
\label{stop}
\mathbb{P}_{\zeta}  =1-\mathbb{P}_{\eta}=\frac{T^{wait}P^{wait}\delta_{int}\nu}{2+T^{wait}P^{wait}\delta_{int}\nu}.
\end{equation} }

 Vehicle $v_p$ as caching provider has a local cache with capacity of  $C_p$. We define $x_{ip}\in\{0,1\}$ as the content caching variable.  If the content of caching requester $v_i$ is cached at caching provider $v_p
 $, $x_{ip} =1$. Otherwise, $x_{ip} =0$. When  $v_i$ caches its content on  $v_p$, it has to make a certain payment for caching resource usage.  The payment that  $v_i$  pays to   $v_p$ is defined as $coin^{v_i\rightarrow v_p} = x_{ip} \varsigma c_{v_i}$, where $ \varsigma>0$ is the price for unit  caching resource.  Since the amount of caching resource on each vehicle is limited, the  total  occupied cache resource of all contents on  $v_p$ cannot exceed its caching capacity, i.e., $\sum_{i\in\mathcal{I}}x_{ip}c_{v_i}\leq C_p$.

Vehicles  can communicate with each other if the distance between them does not exceed the communication distance, i.e., $d_{ip}<\gamma$ \cite{7147772}.   The communication data rate between vehicle $v_i$ and vehicle $v_p$ can be expressed as 
\begin{equation}
\label{rate_V2V}
R_{ip} ={b}\log_{2}(1+\frac{p_ih_{ip}d_{ip}^{-\alpha}}{\sigma^2}), 
\end{equation}
where $b$ is the channel bandwidth, $p_i$ is the transmission power of $v_i$, $h_{ip}$ is the channel gain,  $\alpha$ denotes the path loss exponent, and $\sigma^2$ is the noise power. 

According to (\ref{rate_V2V}), the V2V content transmission latency is
\begin{equation}
\label{delay}
T_{ip} = \frac{c_{v_i}}{R_{ip}}.
\end{equation} 
Since the content of caching requester $v_i$ should be transmitted within $\tau_{v_i}$, we have $\sum_{p\in\mathcal{P}}x_{ip}T_{ip}\leq\tau_{v_i}$.

The total consumed energy consists of two parts: transmission energy consumption and content caching energy consumption. Let $\beta$ denote the price per energy consumption. The energy cost for V2V content caching is
\begin{equation}
E_{ip} = \beta\{p_i\frac{c_{v_i}}{R_{ip}}+e_0*c_{v_i}\},
\end{equation} 
where $e_0$ is the unit energy consumption per caching resource.

Under the constraints of caching capacity and maximum content delivery latency, the problem to maximize system utility is formulated as follows:  
\begin{subequations}
	\label{S1}
	\begin{align}
	\centering
		\max & \sum_{i\in\mathcal{I}}\sum_{p\in\mathcal{P}}( x_{ip} \varsigma c_{v_i}- x_{ip} E_{ip})\notag\\    	
	          &\sum_{i\in\mathcal{I}}x_{ip}c_{v_i}\leq C_p, ~~~\forall p\in\mathcal{P} \label{c1}\\
			&\sum_{p\in\mathcal{P}}x_{ip}T_{ip}\leq\tau_{v_i},~~~\forall i\in\mathcal{I}\label{c2}\\
			&~x_{ip}\in\{0,1\}, ~~~\forall i\in\mathcal{I},~p\in\mathcal{P} \label{c3}
			\end{align}
\end{subequations}
Since $x_{ip}$ is a binary variable, the feasible set and objective function of problem (\ref{S1}) are not convex.  Though we can use an approximate algorithm to solve it,  the scalability of  the solution is very weak as  the solution may fail with the increasing number of vehicles. Moreover, the movement of vehicles leads to time-varying wireless channel such that the conventional optimization method is impractical.  Since deep reinforcement learning is suitable for decision-making problems with high-dimensional and time-varying features, here  we  attempt to utilize it to solve  problem  (\ref{S1}).
\begin{figure*}[!h]
\centering
\includegraphics[height=2.0 in]{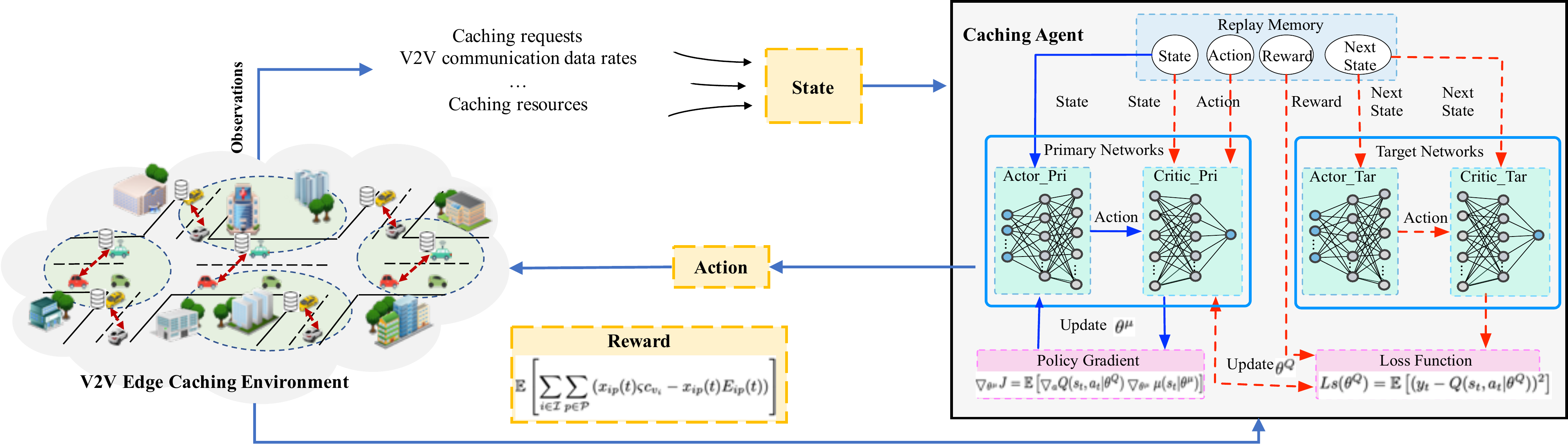}
\caption{DRL-empowered V2V content caching }
\label{DRL_basedV2V Content Caching }
\end{figure*}

\subsection{DRL-based V2V Content Caching Solution}

 We  first reformulate  problem  (\ref{S1}) as deep reinforcement learning form with system state, action, and reward, as shown in Fig. \ref{DRL_basedV2V Content Caching }. Then, we  propose the DRL-based V2V content caching algorithm.

 

System state is  a space to reflect the observed vehicular  environment.  Let $\mathcal{S}$ denote the system state space. The state $s_t\in\mathcal{S}$ at time slot $t$ can be defined as 
\begin{equation}
{s}_t =  \{ R(t), T(t), E(t),\eta(t),F_i,C\},
\end{equation}
where
\begin{itemize}
\item $R(t) =[R_{11}(t),...,R_{IP}(t)]$: is a vector which represents V2V communication data rate between vehicles at time slot $t$;
\item $T(t) =[T_{11}(t),...,T_{IP}(t)]$: is a vector which represents content  latency  via V2V transmission at time slot $t$;
\item $E(t) =[E_{11}(t),...,E_{IP}(t)]$: is a vector which represents energy consumption of V2V content delivery at time slot $t$;
\item $\eta(t) = [\eta_1(t),...,\eta_I(t),\eta_{I+1}(t)..., \eta_{I+P}(t)]$: is a vector which represents each vehicle's driving direction  at time slot $t$;

\item ${F}_i=\{[c_{v_1},t_{v_1}],..,[c_{v_I},t_{v_I}] \}$: is a matrix which represents  the required caching resource and the maximal content delivery latency of  caching requesters;

\item $C= [C_1,..,C_p] $: is a vector which represents  the caching capacities of caching providers.

\end{itemize}
Because of mobility, the location  of each vehicle is time-varying such that V2V communication data rates, content   transmission latency, and   the energy consumption of V2V content delivery are time-varying.

The action of V2V content caching is to match caching pairs.    Let $\mathcal{A}$ denote the action space. The action $a_t\in\mathcal{A}$ at time slot $t$ is defined as 
\begin{equation}
a_t = [x_{11}(t),...,x_{IP}(t)],
\end{equation}
 
After  taking  action $a_t$,  the system will receive an immediate reward $\varUpsilon(s_t,a_t)$. Since the objective of  problem (\ref{S1}) is to maximize system utility, we define the  immediate reward as 

\begin{small}
\begin{equation}
\label{reward1}
\varUpsilon(s_t,a_t)=
\begin{cases}
\mathbb{E}\left[\sum_{i\in\mathcal{I}}\sum_{p\in\mathcal{P}}\left(x_{ip}(t) \varsigma c_{v_i}- x_{ip}(t)E_{ip}(t) \right)\right], \\
~~~~~~~~~~~ \text{if~~(\ref{c1}) and (\ref{c2});} \\
plt, ~~~~~~~\text{otherwise;} \\
\end{cases}
\end{equation}
\end{small}
If the action of caching pairs  satisfies  constraints (\ref{c1}) and (\ref{c2}), the immediate reward is the current system utility. Otherwise, the system will  receive a penalty and $\varUpsilon(s_t,a_t) = plt$, where $plt$ is a negative  constant. 
 The optimal V2V content caching strategy is to maximize the long-term reward which can be defined as 
 \begin{equation}
Reward = \max E\left[\sum_{t=0}^{T-1} \epsilon^t\varUpsilon(s_t,a_t)\right],
 \end{equation}
 where $\epsilon\in[0,1]$ is the discounted factor.

 Based on system state, action, and reward, we attempt to utilize DRL to solve the proposed  V2V content caching problem. There are three common DRL algorithms: Q-learning, Deep Q Network (DQN), and Deep Deterministic Policy Gradient (DDPG). Q-learning is a classical  deep reinforcement learning algorithm which computes the Q-function of each state-action pair for action exploration. However, Q-learning  is not an ideal algorithm for the problem with a high-dimensional observation space. DQN is  a kind of  deep reinforcement learning algorithm which uses deep neural networks instead of Q-function to explore actions.  DQN  is a powerful tool  that can  learn optimal policies with high-dimensional observation spaces but it can only hand low-dimensional action spaces \cite{lillicrap2015continuous}. DDPG is an actor-critic and model-free algorithm that can learn policies in  high-dimensional observation spaces and  high-dimensional action spaces. In this paper, we  exploit the  deep deterministic policy gradient \cite{lillicrap2015continuous}, to solve V2V content caching problem.

 According to DDPG, caching agent is composed of three modules: primary network, target network, and replay memory. Primary network aims to match content caching pairs by policy gradient method.   Primary network  consists of two deep neural networks, namely primary actor neural network and  primary critic neural network.   Target network  is used to generate target value for training primary  network.  The structure of target network is similar to the structure of  primary network but with different parameters.  Replay memory is used to store experience tuples. Experience tuples  include current state, the selected action, reward, and next state, which can be randomly sampled for training primary network and target network.  The detailed interaction processes among these models are shown in Fig. \ref{DRL_basedV2V Content Caching }. 
 
The  explored policy can be defined as a function parametrized  by  $\theta_{\pi}$, mapping current state to  an action $\hat{a} = \pi(s_t|\theta_\pi)$ where  $\hat{a}$ is a proto-actor action generated by the mapping and $\pi(s_t|\theta_\pi)$ is the explored edge caching and content delivery policy produced by primary actor neural network. By adding an Ornstein-Uhlenbeck noise $\mathfrak{N}_t$, the constructed action can be described as \cite{lillicrap2015continuous}
  \begin{equation}
  \label{action}
 a_t = \pi(s_t|\theta_\pi)+\mathfrak{N}_t.
  \end{equation} 
  
  The  primary actor neural network updates network parameter  $\theta_{\pi}$ using the sampled policy gradient, computed as
   \begin{equation}
   \label{pg}
 \bigtriangledown_{\theta_{\pi}}J\approx \mathbb{E} \left[\bigtriangledown_{a}Q(s,a|\theta_{Q})|_{s=s_t,a=\pi(s_t)}   \bigtriangledown_{\theta_{\pi}} \pi(s|\theta_\pi)|_{s=s_t}\right],
   \end{equation}
  where $Q(s,a|\theta_{Q})$ is an action-value function  and will be introduced in the following. Specifically, at each training step, $\theta_{\pi}$ is updated by a mini-batch experience $<s_t,a_t,\mathcal{R}^{imm},s_{t+1}>,$ $ t \in \{1,...,V\}$, randomly sampled from replay memory,
  \begin{equation}
  \small
 \theta_{\pi} = \theta_{\pi}-\frac{\alpha_{\pi}}{V}\sum_{t=1}^{V}\left[\bigtriangledown_{a}Q(s,a|\theta_{Q})|_{s=s_t,a=\pi(s_t)} \bigtriangledown_{\theta_{\pi}} \pi(s|\theta_\pi)|_{s=s_t}\right],
  \end{equation}
  where $\alpha_{\pi}$ is the learning rate of the primary actor neural network.

 The primary critic neural network evaluates the performance of the selected action  based on the action-value function.  The action-value function is calculated by the Bellman optimality equation and can be expressed as 
  \begin{equation}
 Q(s_t,a_t|\theta_{Q}) = \mathbb{E} \left[\mathcal{R}^{imm}(s_t,a_t)+\varepsilon Q(s_{t+1},\pi(s_{t+1})|\theta_{Q}) \right],
  \end{equation}
  Here, the primary critic neural network takes both  current state $s_t$ and next state $s_{t+1}$ as input to calculate $Q(s_t,a_t|\theta_{Q})$  for each action.

The primary critic neural network updates the network parameter  $\theta_{Q}$ by minimizing the loss function $Ls(\theta_{Q})$. The   loss function  is defined as 
  \begin{equation}
  \label{loss}
 Ls(\theta_{Q}) = \mathbb{E} \left[(y_t-Q(s_t,a_t|\theta_{Q}) )^2\right],
  \end{equation}
  where $y_t$ is the target value and can be obtained by
  \begin{equation}
  \label{target}
 y_t = \mathcal{R}^{imm}(s_t,a_t)+\varepsilon Q'(s_{t+1},(\pi)'(s_{t+1}|\theta_{\pi}^T)|\theta_{Q}^T).
  \end{equation}
  where $Q'(s_{t+1},\pi'(s_{t+1}|\theta_{\pi}^T)|\theta_{Q}^T) $ is obtained through the target network, i.e., the network with parameters $\theta_{\pi}^T$ and $\theta_{Q}^T$. 
  
{ The gradient of loss function $Ls(\theta_{Q})$ is calculated by its first derivative, which can be denoted as  \cite{lillicrap2015continuous} 
   \begin{equation}
   \label{loss_gra}
  \bigtriangledown_{\theta_{Q}}Ls = \mathbb{E} \left[2(y_t-Q(s_t,a_t|\theta_{Q}) )\bigtriangledown_{\theta_{Q}}Q(s_t,a_t)\right].
   \end{equation}
   
  According to (\ref{loss_gra}),  the  parameter $\theta_{Q}$ of  primary critic neural network can be updated. Specifically,
  at each training step, $\theta_{Q}$ is updated with a mini-batch experiences $<s_t,a_t,\mathcal{R}^{imm},s_{t+1}>,$ $ t \in \{1,...,V\}$, that randomly sampled from replay memory,
   \begin{equation}
  \theta_{Q} = \theta_{Q}-\frac{\alpha_{Q}}{V}\sum_{t=1}^{V}\left[2(y_t-Q(s_t,a_t|\theta_{Q}) )\bigtriangledown_{\theta_{Q}}Q(s_t,a_t)\right],
   \end{equation}
   where $\alpha_{Q}$ is the learning rate of the primary critic neural network. 
    }
    
  The target network can be regarded as an old version of the primary network with different parameters $\theta_{\pi}^T$ and $\theta_{Q}^T$. At each iteration, the parameters $\theta_{\pi}^T$ and $\theta_{Q}^T$ are updated based on the following definition:
   \begin{equation}
  \label{target_update}
 \begin{split}
 &\theta_{\pi}^T =\omega\theta_{\pi}+(1-\omega)\theta_{\pi}^T, \\
 &\theta_{Q}^T =\omega\theta_{Q}+(1-\omega)\theta_{Q}^T ,
 \end{split}
   \end{equation}
   where $\omega\in [0,1]$.

 DRL-based V2V content caching algorithm is  shown in Algorithm \ref{outer}. First, caching agent  initializes policy $\mu(s|\theta^\mu)$  with parameter $\theta^\mu$ and   initializes  action-value faction  $Q(s,a|\theta^Q)$ with parameter $\theta^Q$. The parameters  of the target network are also initialized. Then, for each time step,  primary  network   generates action $a_t$  based on  current policy $\mu(s|\theta^\mu)$  and current state $s_t$. Observing reward $\varUpsilon(s_t,a_t)$ and next state $s_{t+1}$, caching agent constructs a tuple $<s_t,a_t,\varUpsilon(s_t,a_t),s_{t+1}>$ and stores it into replay memory.  Based on mini-batch technique,   caching agent updates parameter $\theta^Q$   by minimizing  loss function $Ls(\theta^{Q}) $  and   updates $\theta^\mu$ using the sampled policy gradient. The parameters of the target networks are updated based on  $\theta^\mu$, $\theta^Q$, and $\omega$, where $\omega \in [0,1]$. 
 
 \begin{algorithm}[!t]
 
 	\caption{DRL-based V2V content caching algorithm}
 		\label{outer}
 	\begin{algorithmic}[1]
 		\Require  The parameters about mobility model $\delta_{int}$, $\nu$, $T^{wait}$,  and $P^{wait}$ ;
 		                
 		                 The parameters about  V2V communication,  $\gamma$, transmission power, bandwidth, channel gain, and path loss exponent;
 		                 
 		                The state of the observed vehicular environment $s_t$;
 		               
 		\Ensure The explored caching pairs;
 		
   \State Initialize  $\mu(s|\theta^\mu)$ and  $Q(s,a|\theta^Q)$ of the primary network  with parameters $\theta^\mu$ and $\theta^Q$;
 	\State Initialize  the target  network  with parameters $\theta^{\mu'}\leftarrow\theta^{\mu}$ and $\theta^{Q'}\leftarrow\theta^{Q}$;  Initialize replay memory;
     \For {each episode}
 		\State Setup vehicular environment; 
 	 	\For {each time step $t$ }
 	     	\State Execute  action $a_t$ based on  $\mu(s|\theta^\mu)$  and state $s_t$ .
 	        \State Observe reward $\varUpsilon(s_t,a_t)$ and state $s_{t+1}$ based on   (\ref{reward1});
 	        \State Store the tuple $<s_t,a_t,\varUpsilon(s_t,a_t),s_{t+1}>$ into replay memory;
 	        \State Sample a mini-batch of tuples from replay memory;
 	        \State Compute the target value $y_t$ and update $\theta^Q$   by minimizing the loss function (\ref{loss});
 	        \State Update $\mu(s|\theta^\mu)$ using the sampled policy gradient (\ref{pg});
 	        \State Update target networks with:
 	        \NoNumber{$\theta^{\mu'}\leftarrow\omega\theta^{\mu}+(1-\omega)\theta^{\mu'}$}
 	     \NoNumber{$\theta^{Q'}\leftarrow\omega\theta^{Q}+(1-\omega)\theta^{Q'}$}         	         
 		\EndFor 
     \EndFor
 	\end{algorithmic}
 \end{algorithm}

  \subsection{Action Refinement}
 
 {The outputs from DRL-based content caching are continuous values. However,  content caching  variables are integer values, i.e., $x_{ip}\in\{0,1\} $. Therefore, we need to refine the outputs of  DRL. Here, we adopt rounding technique to make action refinement. }  The rounding technique has three steps: 1) find the continuous solution from $a_t$, 2) construct a weighted bipartite graph to establish the relationship between vehicles and BSs, 3) find an integer matching to obtain the integer solution.

 1)  \textit {Find the continuous solution from $a_t$:} We define  the input sets as $\mathbf{z} = [x'_{11},..,x'_{IP}]$, where $ x'_{ip}\in [0,1]$.
 
 2)  \textit {Construct bipartite graph:}  We 
  construct the weighted bipartite graph $\mathcal{G}(\mathcal{I},\mathcal{P},\mathcal{E})$  to establish the relationship between  caching requesters and  providers.   $\mathcal{I}$ represents the   caching requesters  in the network.  $\mathcal{V} = \{v_{ps}:j = ,1,..,P; s= 1,...,P_p\}$, where $P_p= \lceil\sum_{i =1}^{I}x_{ip}\rceil$  implies   caching  provider $p$ can serve the number of $P_p$  caching requesters. The nodes $\{v_{ps}:s = 1,..,P_p\}$ correspond to caching  provider $p$. The most important procedure for constructing graph $\mathcal{G}$ is  to set the edges and the edge weight between $\mathcal{I}$ and $\mathcal{P}$. The edges in $\mathcal{G}$ are constructed using  Algorithm \ref{edge}.

    
   3)  \textit {Action refinement:}  We utilize the Hungarian algorithm \cite{kuhn1955hungarian} to find a complete max-weighted 
   bipartite  matching $M_{match}$.  According to the  $M_{match}$, we obtain the detailed caching pairs matching. Specifically, if $(i,v_{ps},e_{ips} )$ is in the $M_{match}$, we set $x_{ip}= 1$; otherwise, $x_{ip}= 0$. 
   
 {The outputs of DRL-based content caching are  fractional solutions, the rounding results are  integer solutions.}

   \begin{algorithm}[!tbp]
         	\caption{ Construct the edges of  bipartite graph $\mathcal{G}$}
         	\label{edge}
         	\begin{algorithmic}[1] 
         	\Require  The outputs of the explored caching pairs;
         	 		               
         	\Ensure The constructed bipartite graph $\mathcal{G}(\mathcal{I},\mathcal{P},\mathcal{E})$ ;
         	\State Set $\mathcal{E}\leftarrow\varnothing$.
         	\If{$P_p\leqslant 1$ }
         	        \State There is only one node $v_{p1}$ corresponding to caching  provider $p$.
                   	\For{ each $x'_{ip}>0$}
                         	\State Add edge $(i,v_{p1})$ into edge set $\mathcal{E}$ and let the edge weight as { $e_{ip1} =  x'_{ip}$. }
                 	\EndFor
         	\Else
         	        \State Find the minimum index $i_s$ where  $\sum_{i'=1}^{i_s}x'_{ip}\geqslant s$.
         	        \If{$i = i_{s-1}+1,..,i_{s}-1$, and $x'_{ip}>0$}
         	               \State Add  edge $(i,v_{ps})$  into  edge set $\mathcal{E}$ with weight { $e_{ips} =  x'_{ip}$. }
         	        \ElsIf{ $i = i_s$}
         	             \State Add edge $(i,v_{ps})$  into edge set $\mathcal{E}$ with  $e_{ips} = 1-$$\sum_{i=1}^{i_s-1}x'_{ip}$. 
         	        \Else
         	              \State Add edge $(i,v_{p(s+1)})$  into  edge set  $\mathcal{E}$ with weight 
         	              \NoNumber{$e_{ip(s+1)} =  \sum_{i=1}^{i_s}x'_{ip}-s$.  }
         	      	\EndIf
         	\EndIf
         	\end{algorithmic}  
         \end{algorithm}

\section{Proof-of-Utility Consensus in Vehicular Networks }
\label{blockchain}

In this section, we present the details of PoU consensus mechanism for permissioned blockchain in the edge plane and propose how to evaluate BS utility for block verifier selection.

\subsection{PoU Consensus}

DPoS is a fast and efficient  blockchain consensus mechanism which leverages voting and selection  to protect blockchain from centralization and malicious usage \cite{larimer2014delegated}.   Compared to PoW and PoS, the number of entities participating in DPoS consensus is very small, making it possible to effectively reduce  the time consumption and energy consumption to reach consensus. Inspired by DPoS, the proposed PoU consensus consists of two parts: 1) delegate selection, 2) block production and verification, as shown in Fig.\ref{Block}. Different from DPoS, we are not use the stake of users but use the utility of base station to select delegates. {The details about PoU consensus are introduced in the following.}


1) \textit{Delegate Selection:}
Since coin transfer in content caching transactions occurs  among vehicles and these vehicles are non-trusted, we define vehicles as token holders to dominate delegate selection process. Because delegates involve the key process of consensus algorithms (i.e., block production and validation), they are preferably neutral nodes \cite{larimer2014delegated}. In content caching process,  BSs are not directly participating in content delivery and coin payment, which means they do not  get any profit from the caching process, such that they are ideal neutral nodes and can be regarded as delegate candidates. 

At each  selection, vehicles vote for their preferred BSs with the highest utility. The utility is utilized to measure the quality of BSs.  Higher utility means BS is equppied with more powerful computing and processing abilities to generate and verify block. The details of utility evaluation is given in the following subsection \ref{satis}. Each vehicle has one vote per round and  the voting weight is proportional to the number of coins it holds. The top $\widehat{n}$ candidates with the most votes are selected to form a delegate commission, where $\widehat{n}$  is an odd integer and no greater than the number of BSs.

\begin{figure}
\centering
\includegraphics[width=3.4 in]{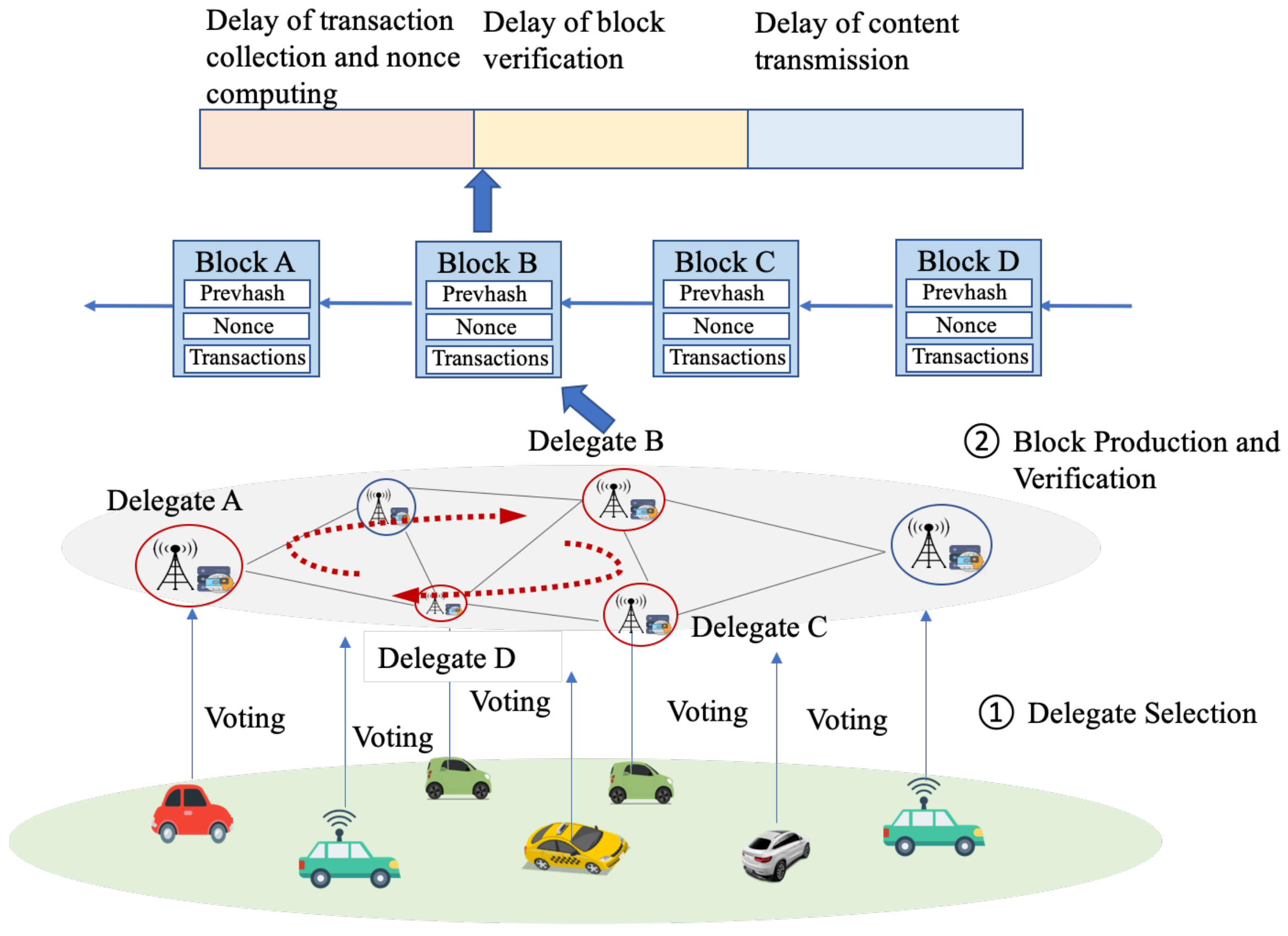}
\caption{The PoU consensus of vehicular permissioned blockchain}
\label{Block}
\end{figure}
2) \textit{Block Production and Verification:} 
In the block production and  verification process, delegates are divided into two roles: leader and verifier, where leader is responsible for transaction collection and block production, and verifier is responsible for  block verification. At each  block production process, one of $\widehat{n}$  delegates acts as the leader and the other delegates act as verifiers. Leader is generated in a round-robin manner among  delegates which indicates each delegate can become a leader to produce block.   For example, in Fig. \ref{Block}, delegate B is the leader responsible for creating Block B. 

In a specific block production and verification process, the leader first collects a certain amount of V2V content caching transactions and then calculate a correct hash to create an unverified block.  The block verification is a three-phase protocol consisting of block broadcast, block verification, and confirm. In block broadcast phase, the leader broadcasts $\widehat{n}-1$ broadcasting messages to other delegates. The broadcasting message has the form: $Bro = \left\langle bro_{msg}||PK_{led}||PK_{ver}||ts_{bro}||block \right\rangle$, where $PK_{led}$ is the source address of the message, $PK_{del}$ is the destination address of the message, $ts_{bro}$ is the time stamp, $block$ is the created block. In the block verification phase, each verifier first verifies the signature of the received broadcasting message.  Then verifiers  audit the correctness of  V2V content caching requests packaged in the newly block  and  broadcast their audit results with their signatures to each other in a distributed manner. In the confirm phase, each  verifier compares its  audit  result with the received audit results  from other  verifiers and sends a confirm message  to the leader. The confirm message has the form: $Con =  \left\langle con_{msg}||PK_{ver}||PK_{led}||Aud_{self}||Aud_{rec}||Rsu_{comp}\right\rangle$, where $Aud_{self}$ is  the audit result of the verifier own,  $Aud_{rec}$ is the  records of received audit results of other verifiers,  $Rsu_{comp}$ is the comparison result. After receiving all delegates' confirm messages, the leader analyses them and decides the correctness of the block. If more than two third of verifiers agree on the block, the leader will send it  to all delegates to store. The BSs which are not in the delegate commission  will synchronize  the latest blockchain from nearby delegates periodically.

Once the newly produced block has been successfully appended to the blockchain, the BSs participated in block production and verification process will be rewarded to compensate for their resource consumption. If all delegates become the  leader once, the order of the delegates are shuffled and then they  produce the future blocks in a round-robin manner again. If a delegate fails to create a block during its turn, the block is skipped and the transactions in the  skipped block will be transferred to the next one.

\subsection{Utility Evaluation}
\label{satis}
There are $M$ BSs distributed in vehicular networks, denoted as $\mathcal{M} = \{1,...,M\}$. All $M$ BSs can communicate with each other at the speed of $r$ via wired line. In the delegate selection process, $\widehat{n}$ of $M$ BSs  are elected by vehicles to group as a delegate commission, denoted as $\widehat{\mathcal{N}} = \{1,...,\widehat{n}\}$.  From the perspective of vehicles, they prefer to vote for the BSs providing fastest block production and verification as  delegates. In this paper, we adopt utility function with respect to  time consumption to evaluate the  performance of BSs. 
 
A short time consumption leads to a good user experience thus the utility function should monotonically decrease with time consumption. To satisfy content delivery constraint,  the utility is set as $0$ if any content caching time consumption of vehicles is exceeded its maximal content delivery latency $\tau_{v_i}$.  We consider  that $K$ transactions are collected in the $k$-th block and ${\tau^k} = \min\{{\tau_1},{\tau_2},...,{\tau_K}\}$, where $K\leq I $. 
The utility function of BS $m$ is defined as:
   \begin{equation}
 U_m^k=\left[e^{1-{T_m^k }/{\tau^k}}-1\right]^+,
   \end{equation}
   where $T_m^k$ is the  time consumption of block production and verification on BS $m$ and $[x]^+ = \max\{x,0\}$.
 $T_m^k$ is consisted of three parts: 1)  the delay of transaction collection and hash computing, 2) the delay of block verification, 3) the delay of content transmission, which can be described as
  \begin{equation}
T_m^k = T_{k}^H+ T_{k}^V +T_{ip}.
  \end{equation}
According to DPoS,  the delay of  transaction collection and hash computing $T_{k}^H$ is pre-defined, such as 0.5s in EOS. The content transmission $T_{ip}$ is shown in Eq. (\ref{delay}).

 The  delay of block verification  consists of three parts: 1) block broadcasting, 2)  cross-verification among verifiers, 3) block confirm, which can be described as:
    \begin{equation}
    \label{bv}
   T_{k}^V = T_k^{bb}+ T_k^{cv} +T_k^{bc}.
    \end{equation}
BS $j \in \widehat{\mathcal{N}}$ is the leader to dominate  block verification process.  The other BSs are  verifiers to audit the produced block, denoted as  ${j'} \in \widehat{\mathcal{N}} /\{j\}$.  Since the leader $j$  broadcasts its produced block to verifiers simultaneously, the block broadcasting time is determined by the longest block transmission time, thus
  \begin{equation}
T_k^{bb} =  \max_{{j'} \in \widehat{\mathcal{N}} /\{j\}}\{ \dfrac{I_kd_{jj'}}{r}\} ,
  \end{equation}
  where $d_{jj'}$ is the distance between the leader $j$  and verifier $j'$, $I_k$ is the size of the $k$-th  block before verification.
 Cross-verification consists of  three steps. Each verifier first performs  local verification to  verify the raw block  from the leader and then broadcasts its local-verified result  to other verifiers. After receiving the local-verified result, verifiers performs  second audit. We denote $O_k$ as the size of  local-verified result and $W_k$ as the size of  second-audit result. The $k$-th  block  cross-verification  time consumption  can be written as 
 \begin{equation}
T_k^{cv} =  \max_{{j',j''} \in \widehat{\mathcal{N}} /\{j\},j' \neq j''} \{\dfrac{I_kf_0}{F_{j'}}+ \dfrac{O_kd_{j'j''}}{r} +  \dfrac{O_kf_0}{F_{j''}} \},
 \end{equation}
 where $f_0$ denotes the  computation resource for verifying one bit of $I_k$, $F_{j'}$ and $F_{j'i}$ denote the computation resource that verifier $j'$ and verifier $j''$  provide to block verification respectively, and $d_{j'j''}$ denote the distance between  verifier $j'$  and verifier $j''$. 
 Block confirm time is determined by the longest second-audit result transmission time, which is
  \begin{equation}
T_k^{bc} =  \max_{{j'} \in \widehat{\mathcal{N}} /\{j\}}\{ \dfrac{W_kd_{jj'}}{r}\}.
  \end{equation}

Base on Eq. (\ref{bv}), the $T_m^k$  and $ U_m^k$ can be obtained, respectively.   Then, each vehicle  casts it vote to the BS $m^*$, which satisfies 
\begin{equation}
m^* = \arg\max\{U_m^k\},~m\in\mathcal{M}.
\end{equation}
The vote weight is equal to  content caching transaction fee, i.e., $coin^{v_i->v_j}$.

\section{Security Analysis and Numerical Results}
\label{result}
In this section, we first provide security about our proposed  blockchain-based content caching. Then, we evaluate the performance of the proposed V2V content caching scheme based on Uber  dataset  \cite{nyctaxi} and analyse the performance of  the proposed PoU.
\subsection{Security Analysis}

{The use of permissioned blockchain establishes a secure content caching for multiple vehicles without mutual trust.

1)\textit{ Without reliance on a single trusted third party:} The permissioned blockchain reduces the reliance on a trusted curator.   If V2V content caching requires the involvement of a trusted third party, the system security largely depends on the security of the trusted third party. If the centralized security cannot be guaranteed, content in the system faces high risk of  leakage. In our schemes,  caching requesters and caching providers deliver content  a P2P manner without a third party, which makes system robust and scalable.

 { 2)\textit{ Privacy protection:}  
All vehicles and BSs transmit messages about  caching request and blockchain  in a pseudonymous  manner.  Specifically, vehicle $v_i$ uses its public key $PK_{v_i}$ as the pseudonym to guarantee the anonymity of its real identity. 
The messages (i.e., $Req^{v_i\rightarrow b_j} $, $Mes^{v_p\rightarrow b_j} $, $ Resp_{req}^{b_j\rightarrow v_i}$, and $ Resp_{pro}^{b_j\rightarrow v_p}$), and transactions (i.e., $Trans^{v_i\rightarrow b_j} $) are signed and can only be accessed by a specified vehicle with the right private key. If a malicious vehicle wants to forge the signature of vehicle $v_i$ to pass the authentication process, the adversary  has to forge a signature $Sig_{v_i} = Sign_{SK_{v_i}} (\cdot)$. However, adversaries have no access to the private key $SK_{v_i}$. The only information that the adversary can obtain is the public key of vehicle $v_i$. Since there is no feasible solution to obtain the private key from the public key, the adversary  cannot forge the signature information of a legal vehicle. The anonymity and digital signature can protect the privacy of vehicles in V2V content caching. }

{ 3)\textit{ Majority attack:} Majority attack is a  noteworthy security issue where an entity or user can take control of the system and use it for self benefit if the attack is performed properly. In the proposed scheme, all  blocks and transactions are publicly audited and mutual verified by all  elected delegates.    The delegate selection process is democratic where delegates are individually selected by vehicles based on the proposed PoU consensus. The selected delegates form a commission to produce block in a round-robin way. Therefore, it is nearly impossible for an adversary to dominate the delegate selection process and make the majority attack.
}

  4)\textit{ Transaction traceable:}  All broadcasted transactions in blockchain are forever  recorded with a timestamp  and  these transactions cannot be modified by a single entity. Since blockchain is a distributed ledger, transactions are  synchronized updated and can be easily obtained from any BS. When malicious behavior occurs, any vehicle can easily verifies and traces  previous records through accessing a BS. Timestamps in blockchain can be used to keep  transactions  intact, thus leaving no chance to counterfeits.}


\subsection{Permance Analysis of Vehicular Content Caching Scheme }

We use Python and TensorFlow to evaluate the performance of the proposed DRL empowered content caching scheme based on a real-world  dataset  from Uber\cite{nyctaxi}. 

 { The trajectory of Uber dataset is used  to simulate the changing locations of vehicles.} This  dataset has 4.5 million Uber pickups in New York City from April to September 2014, and 14.3 million Uber pickups from January to June 2015, as shown in Fig. \ref{fig4}.  We take  100 vehicles  as examples from an observation area, whose latitude is from 40.668671 to 40.678719, and the longitude is from -73.930269 to -73.950915.  The observed area is approximately 1.52 $km^2$. 

 The data size of each content, required caching resource and the maximal content delivery latency of each content are within the range of $[10, 50]$ MB, $[0.5, 2.5]$ GB, and $[5,10]$s, respectively.
 The caching capacity of vehicles is $5$ GB. The  maximal  distance of V2V communication  is $\gamma=500$ m. The transmission power of vehicles for  content delivery is $24$ dBm. The channel bandwidth  is $10$ MHz.  
The noise power is $\sigma^2 =10^{-11} $ mW. The proposed DRL empowered algorithm is  deployed on a MacBook Pro laptop, powered by two Intel Core i5 processor (clocked at 2.6Ghz).  The activation function  is $\frac{tanh(x)+1}{2}$.  The maximum episode is 4000 and the maximum number of steps in each episode is set to 20. The  size  of mini-batch is  set up as $32$ .  The penalty is -100. 

   \begin{figure}
   	\centering
   	\includegraphics[width =3.3 in]{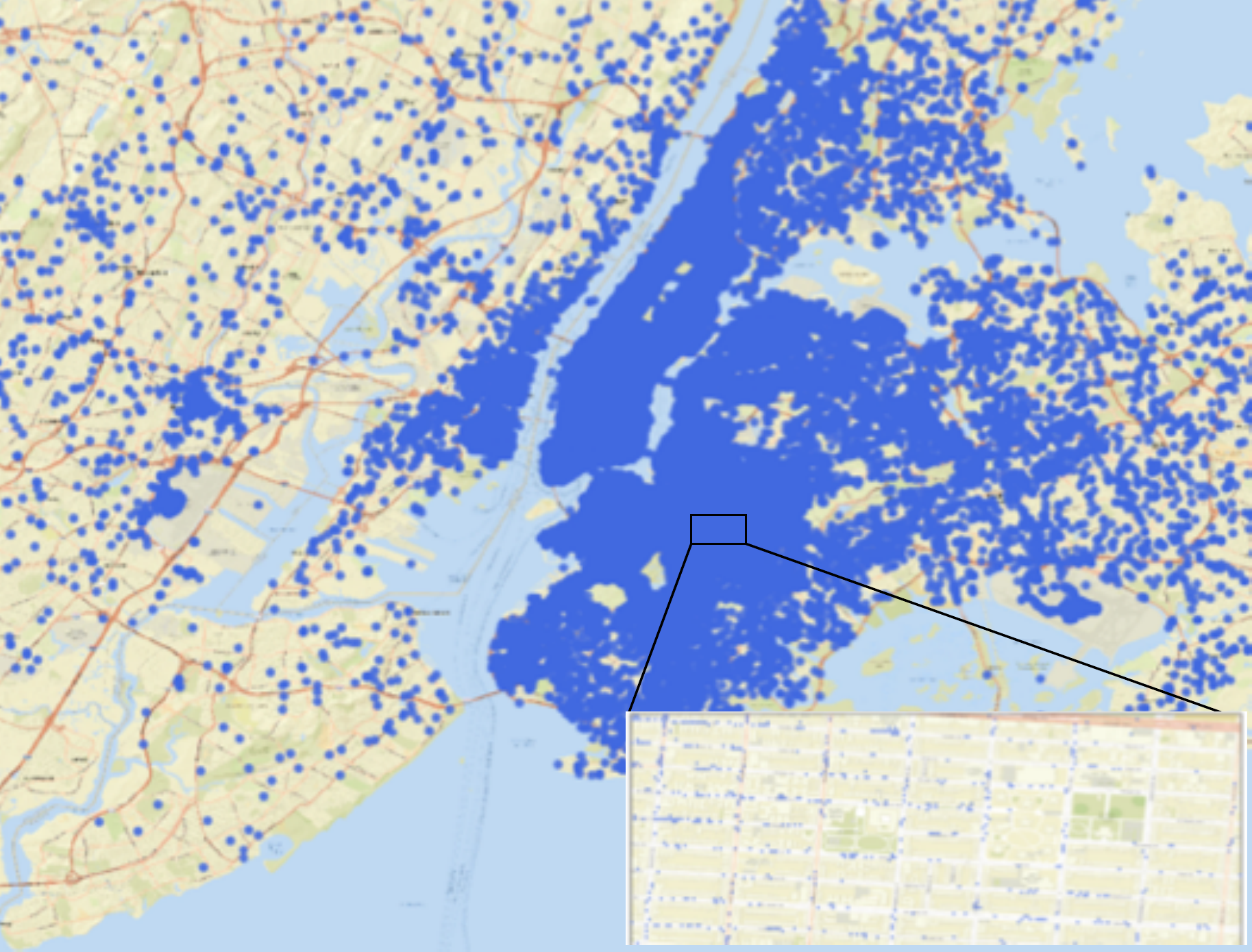}
   	\caption{Spatial distribution of vehicle trace points.}
   	\label{fig4}
   \end{figure}
   
To verify the performance of our proposed DRL empowered  edge caching and content delivery  algorithm, we introduce the following two benchmark schemes,
\begin{itemize}

\item \textit{Greedy content caching (GCC)}: In this scheme,
each caching requester delivers its content to the caching provider with the highest wireless communication data rate.

\item \textit{Random content caching (RCC) }:  This scheme randomly selects a caching provider  for a caching requester to perform content caching. Note that the distance between the caching provider and the caching requester should not exceed $\gamma$.

\end{itemize}
 \begin{figure}
      	\centering
      	\includegraphics[width =3.5in]{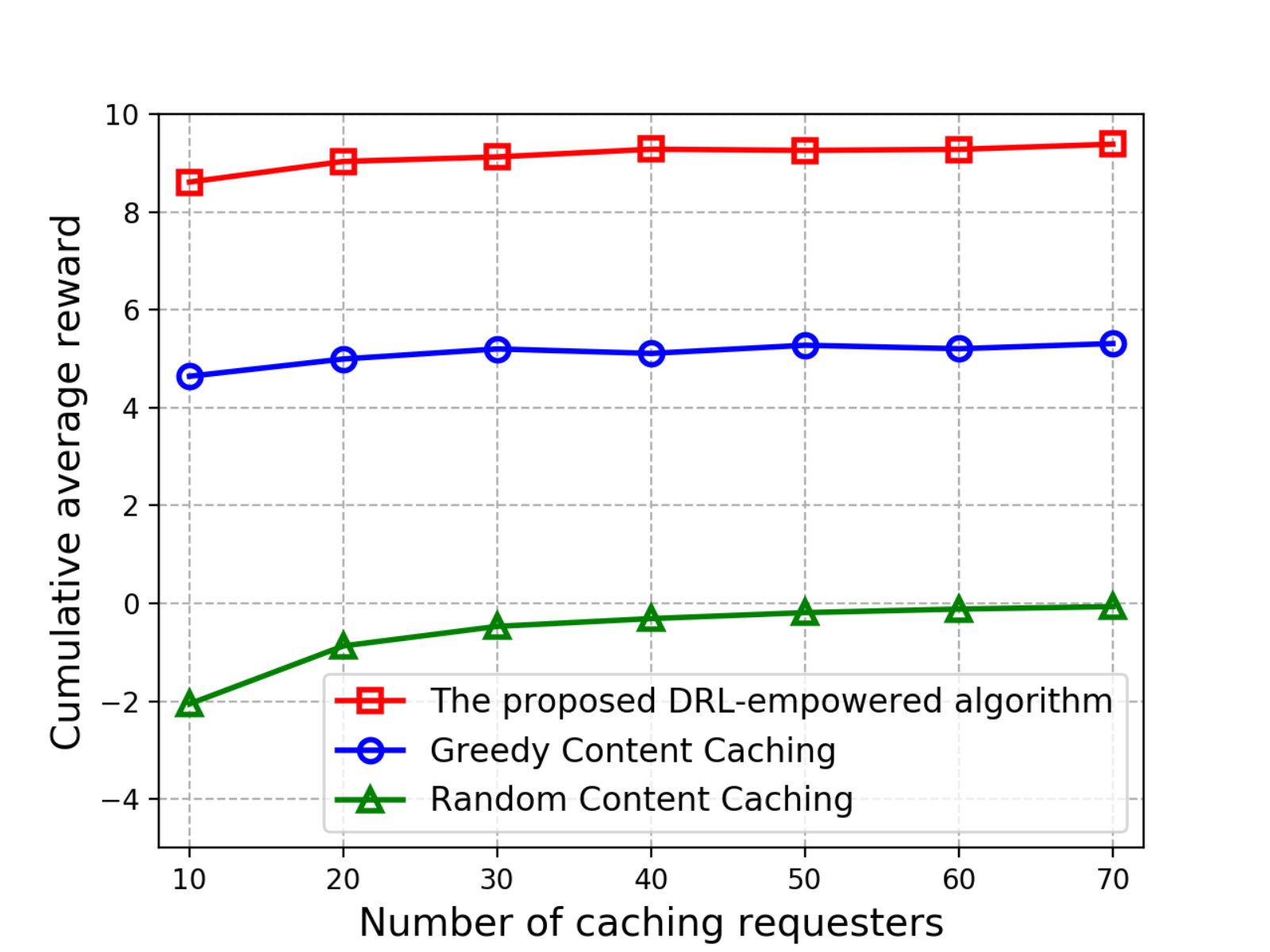}
      	\caption{Comparison of cumulative average  reward with respect to number of caching requesters under different scheme.}
      	\label{fig7}
 \end{figure}

 We set the number of caching provider as 50.  Fig. \ref{fig7} plots the comparison of cumulative average reward with respect to number of caching requesters under different scheme. From Fig.  \ref{fig7} we can draw several observations. First, the performance of the proposed DRL-empowered algorithm significantly outperforms the two benchmark policies. The reason is that  the proposed DRL-empowered algorithm designs the content caching policy based on current network topology  and wireless channel condition while GCC and RCC are not able to acquire real-time parameters of the vehicular network, such as available caching resource of caching providers. Second,  the performance of GCC is better than that of RCC because of taking  wireless communication data rate into account. Third,  the cumulative average reward of RCC is the lowest, even lower than $0$ (i.e.,  receiving a penalty). This implies that  randomly choosing a caching provider  for a caching requester  results in unsuccessful content caching.
     \begin{figure}
            	\centering
            	\includegraphics[width =3.5in]{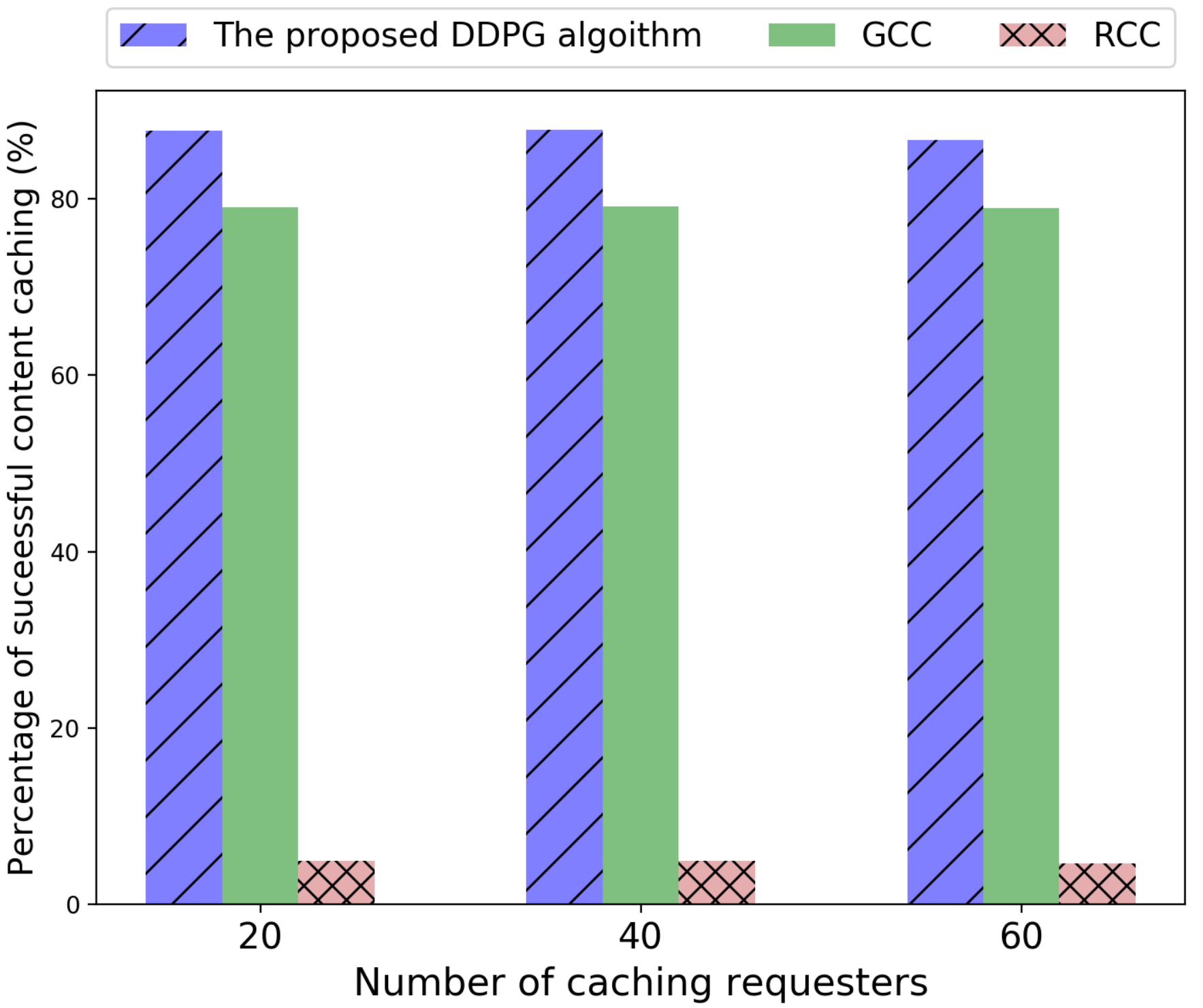}
            	\caption{Comparison of percentage of successful content caching with respect to number of caching requesters under different scheme.}
            	\label{fig8}
    \end{figure}

 Fig. \ref{fig8} plots the comparison of  percentage of successful content caching with respect to number of caching requesters under different scheme.  We can see that the proposed DRL-empowered algorithm  has a good efficiency that over 86\% caching requesters can successfully execute content caching within their stringent deadline constraints. The performance of GCC is also well that about 78\%  caching requesters can successfully execute content caching.   The performance of RCC is quite poor that only 5\% caching requesters can successfully execute content caching. Thus, we can conclude the proposed algorithm is the most efficient. 
 
    \begin{figure}
      	\centering
      	\includegraphics[width =3.5in]{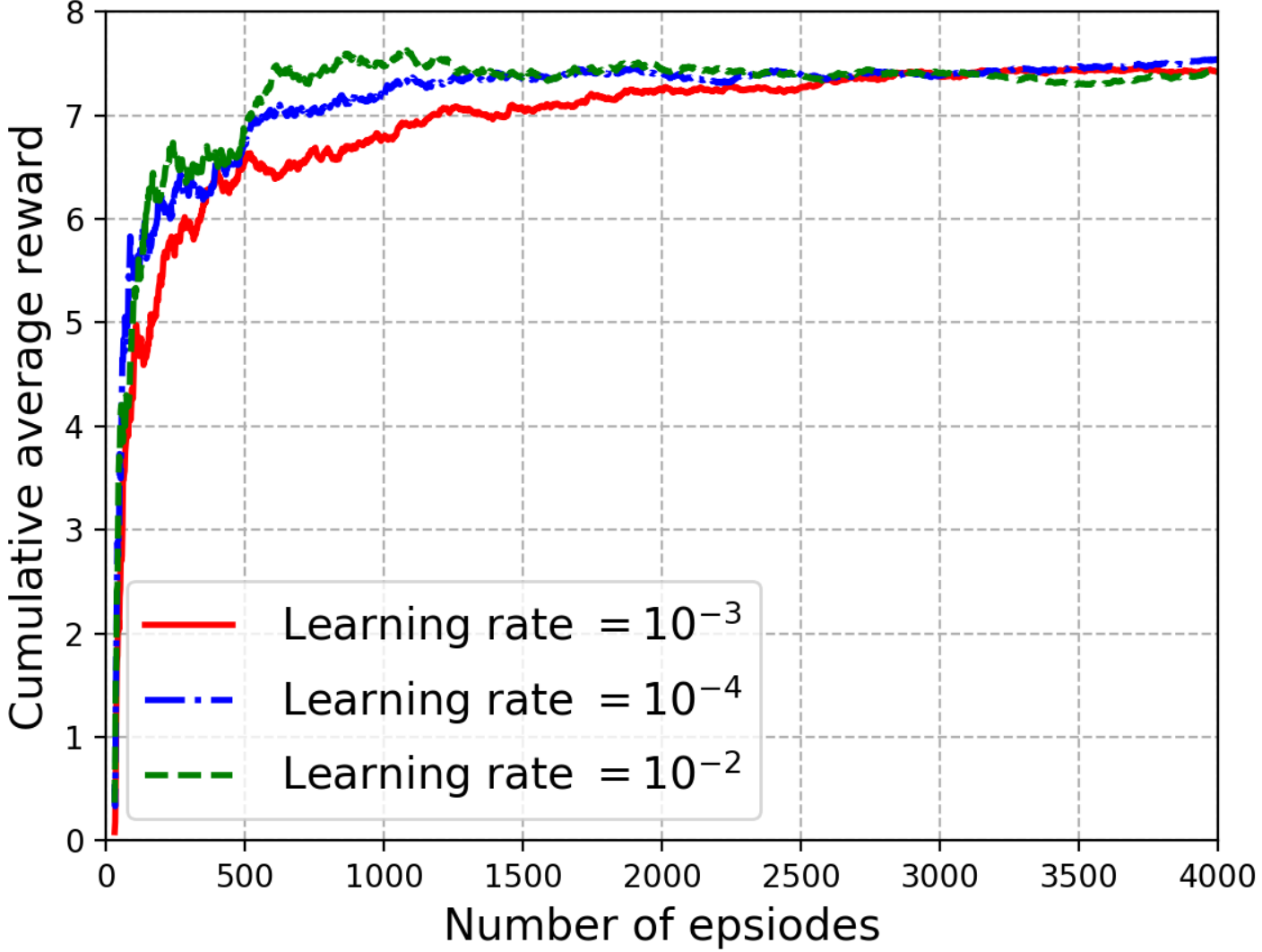}
      	\caption{Impact of  learning rate on the performance of the proposed DRL-empowered  algorithm.}
      	\label{fig5}
      \end{figure}

   \begin{figure}
   	\centering
   	\includegraphics[width =3.5in]{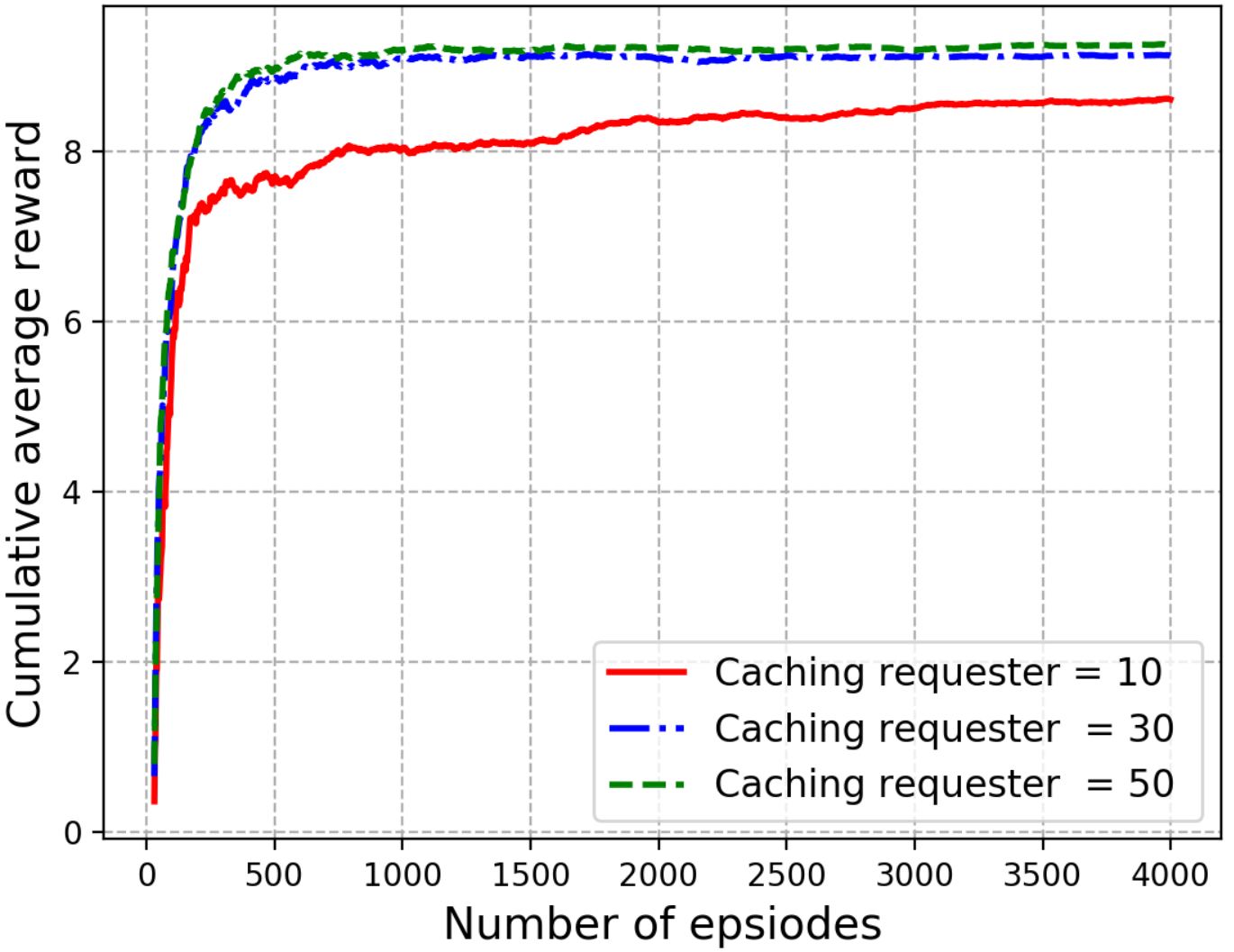}
   	\caption{Impact of number of vehicles on the performance of the proposed DRL-empowered  algorithm.}
   	\label{fig6}
   \end{figure}

 Fig. \ref{fig5}  and Fig. \ref{fig6}  show the impact of  learning rate and the impact of number of vehicles on the performance of the proposed  algorithm. From Fig. \ref{fig5}, we can observe that at all learning rates, the cumulative average rewards converge. When the learning rate is $10^{-2}$,  the proposed algorithm converge  slightly faster than the case when the learning rate is $10^{-3}$ and $10^{-4}$ .  In the proposed algorithm, we set the learning rate as $10^{-2}$.  { From  Fig. \ref{fig6}, we can see that the increasement of  caching requesters results in a high reward, which means the caching utility is improved. Moreover, the reward is greatly increased as the number of caching requesters changes from 10 to 30 but it is slightly increased as the number of caching requesters changes from 30 to 50.   This is reasonable because in all cases the total caching resource of the caching providers is the same, such that when the number is 30, the system utility is already approached to upper bound and the further increasement of caching requesters cannot make a greatly utility improvement.}

 \begin{figure}
    \centering
    \subfigure[]{
    \includegraphics[width =3.3 in]{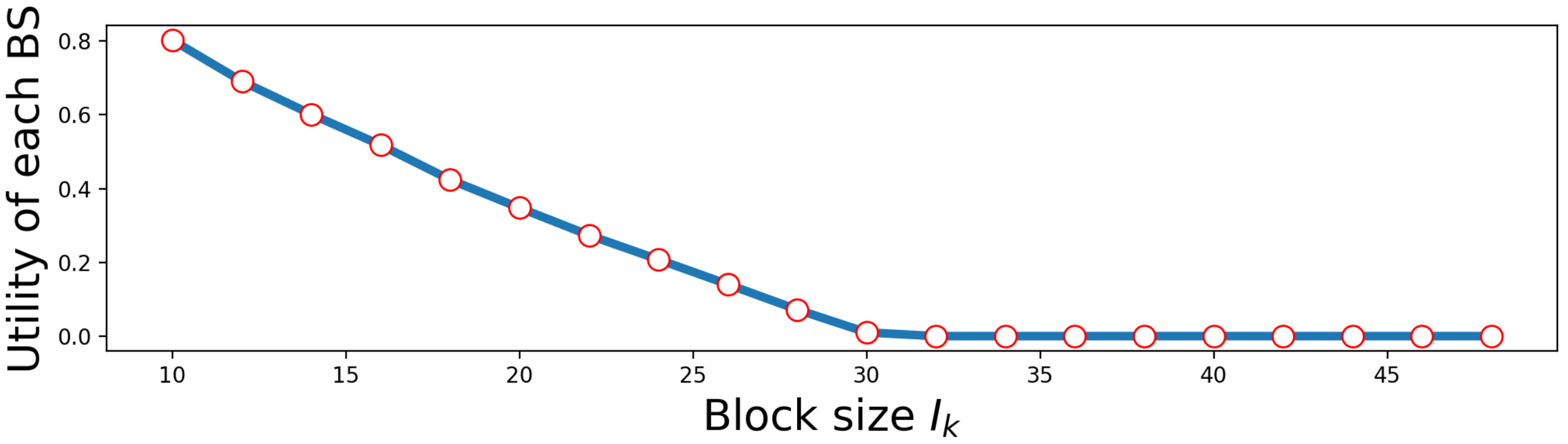}
    }
   	\subfigure[]{
   	 \includegraphics[width =3.3 in]{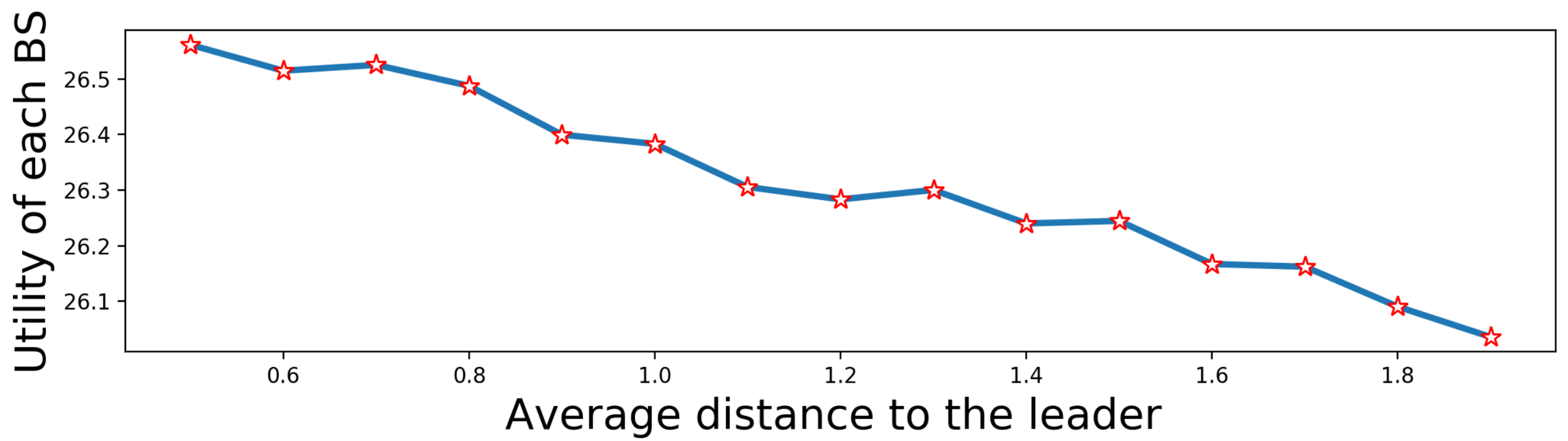}
   	 }
   	  \caption{(a) Utility of base station v.s. Block size $I_k$. (b) Utility of base station v.s. Distance to the leader. }
    \label{fig9}
   \end{figure}

\subsection{Performance Analysis of PoU Consensus}
The size of block before verification $I_k$,  the size of local-verified result $O_k$, and the size of second-audit result $W_k$ are uniform distributed in $[10,50] $ MB, $[1,5]$ MB, and $[100, 500]$ KB, respectively. The computation resources of each BS are within the range of $[5,10]$ GHz.  We evaluate the utility of vehicle $v_i$ towards the BS with various block size in Fig.  \ref{fig9}a.  
From Fig.  \ref{fig9}a, we can see that the utility decreases with the increasement of $I_k$. The reason is that with the increasing in block size, more time consumption is needed to verify the candidate block.  When $I_k$ exceeds  $30$ MB, the utility turns to be 0. This is because the time consumption on blockchain-based content caching exceeds the maximal content delivery latency  (i.e., $T_{v_i}>\tau_{v_i}$). Fig. \ref{fig9}b depicts the utility  of each vehicle with respect to average distance between the vehicle and the PoU selected leader. In general, the utility  decreases with the increasement of distance.  This is because a larger distance  between a vehicle and the PoU selected leader, the more communication time it needed for block verification. We can conclude that the utility decreases with  the increasement of  block size $I_k$ and communication distance, which is in accordance with our previous analysis in subsection \ref{satis}.

\section{Conclusion}
\label{c}
In this article, we have proposed a secure and intelligent content caching for vehicles by integrating deep reinforcement  learning and permissioned  blockchain  in vehicular edge computing  networks.  We first proposed a  blockchain empowered distributed and secure content caching framework  where vehicles acted as caching requesters and caching providers to perform  content caching and BSs acted as verifiers to build and maintain permissioned blockchain. To learn dynamic  network topology and time-variant wireless channel condition,  we utilized DRL to design an optimal content caching scheme.  We exploited permissioned blockchain  to maintain the security and privacy of content caching among vehicles and introduced a new verifiers selection metric, PoU to accelerate  block verification.   Security analysis shows that our proposed blockchain empowered content caching can achieve security and privacy protection. Numerical results based on a real dataset from Uber indicate that the DRL-inspired content caching scheme significantly outperforms two benchmark policies.

\bibliographystyle{IEEEtran}
\bibliography{reference}
\end{document}